\documentclass[journal]{IEEEtran}
\usepackage{cite,amsmath,graphicx,bbm}
\usepackage{algorithm}
\usepackage{subcaption}
\usepackage{caption}
\usepackage{url}
\usepackage{booktabs}
\usepackage{algpseudocode}
\usepackage{hyperref}
\usepackage{multicol}
\usepackage[table]{xcolor}

\usepackage{letltxmacro}
\LetLtxMacro{\originaleqref}{\eqref}
\renewcommand{\eqref}{Eq.~\originaleqref}

\usepackage{color}
\IEEEpubid{ 
\begin{minipage}{\textwidth}\ \\ \tiny
Copyright \copyright~2019 IEEE. Personal use of this material is permitted. However, permission to use this material for any \\
other purposes must be obtained from the IEEE by sending an email to pubs-permissions@ieee.org.
\end{minipage}
}

\title{Low Power Depth Estimation of Rigid Objects for Time-of-Flight Imaging}

\author{James~Noraky,~\IEEEmembership{Student Member,~IEEE,}
	Vivienne~Sze,~\IEEEmembership{Senior Member,~IEEE}}

\begin{document}

\maketitle

\begin{abstract}
Depth sensing is useful in a variety of applications that range from augmented reality to robotics. Time-of-flight (TOF) cameras are appealing because they obtain dense depth measurements with minimal latency. However, for many battery-powered devices, the illumination source of a TOF camera is power hungry and can limit the battery life of the device. To address this issue, we present an algorithm that lowers the power for depth sensing by reducing the usage of the TOF camera and estimating depth maps using concurrently collected images. Our technique also adaptively controls the TOF camera and enables it when an accurate depth map cannot be estimated. To ensure that the overall system power for depth sensing is reduced, we design our algorithm to run on a low power embedded platform, where it outputs $640\times480$ depth maps at 30 frames per second. We evaluate our approach on several RGB-D datasets, where it produces depth maps with an overall mean relative error of 0.96\% and reduces the usage of the TOF camera by 85\%. When used with commercial TOF cameras, we estimate that our algorithm can lower the total power for depth sensing by up to 73\%.
	
\end{abstract}


\section{Introduction}
\label{sec:intro}

Depth sensing is useful in a variety of applications that range from augmented reality to robotic navigation. One common way to measure depth is to use a time-of-flight (TOF) camera. TOF cameras obtain depth by emitting light and measuring its round-trip time. Compared to other depth sensors, TOF cameras are appealing because they are compact, have no moving parts, and obtain dense depth measurements with minimal computation and latency \cite{Hansard2013}. The depth measurements obtained by a TOF camera are represented as a depth map, which is an image whose pixel values represent the distance from the sensor to different points in the scene. 

\begin{figure}[h]
	\begin{center}
	\includegraphics[scale=0.33]{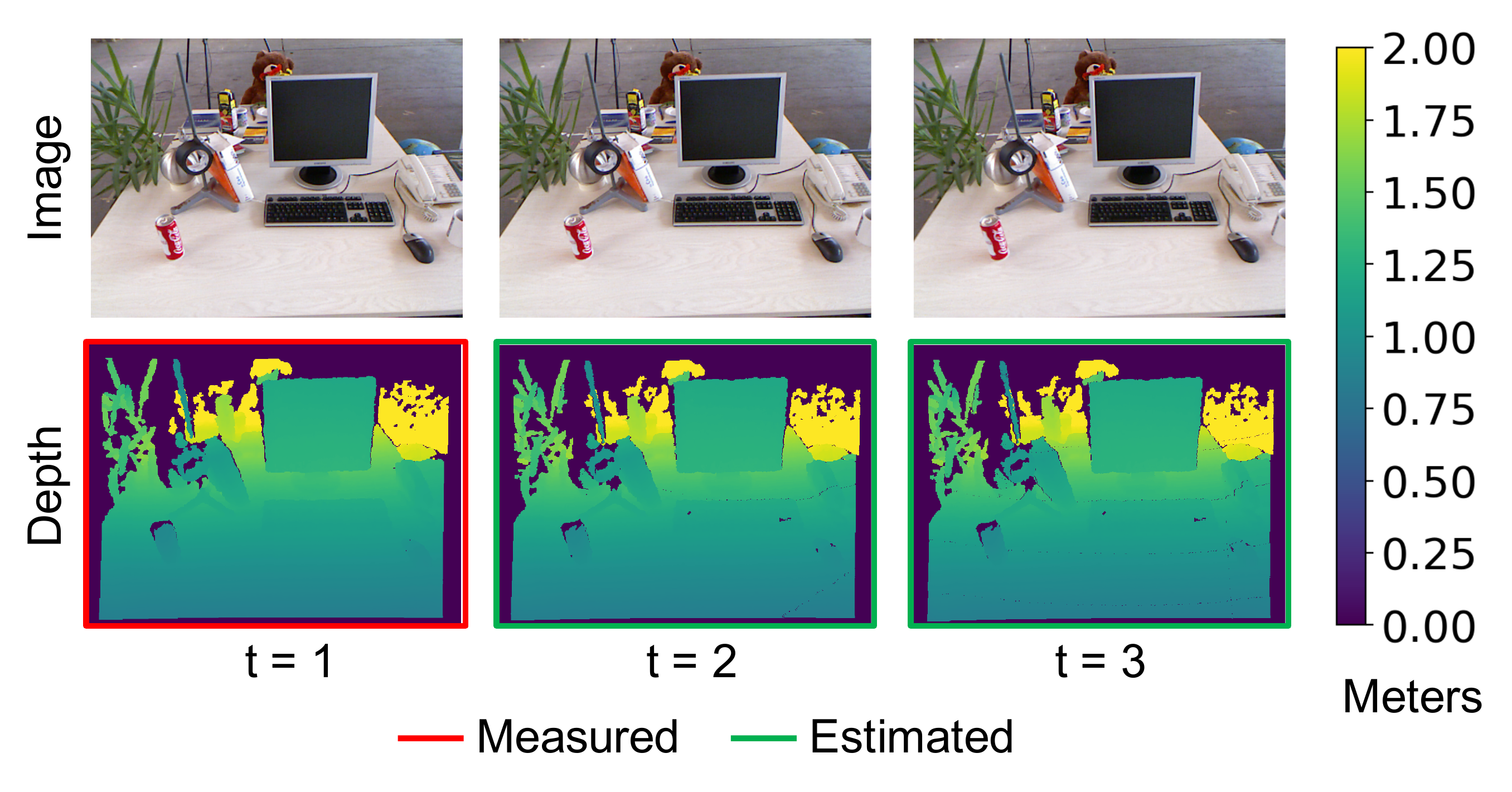}
	\end{center}
	\caption{\textbf{Depth Estimation Setup}: We estimate causal depth maps using concurrently collected images and previously measured depth. The TOF camera is used when an accurate depth map cannot be estimated.}
	\label{fig:setup}
\end{figure}

However, for applications that run on mobile devices, one drawback of using a TOF camera is that its illumination source is often power hungry, where continuous acquisition of depth can limit the battery life of the mobile device. One way to address this issue is to reduce the usage of the TOF camera, but this is problematic for applications that require depth in real time, or 30 frames per second. Here, we propose an algorithm to address this issue by estimating depth maps without using the TOF camera as shown in Figure \ref{fig:setup}.

Our technique estimates depth maps by using the pixel wise motion of concurrently collected images, or the optical flow, to estimate changes in the scene and update a previously measured depth map. For many applications, images are routinely collected, and our goal is to reuse them to obtain depth. We focus on estimating the depth of rigid objects and environments, and we show that it is possible to estimate accurate depth maps while significantly reducing the usage of the TOF camera in these scenarios. While the assumption of rigidity may seem restrictive, our approach requires only the local environment that the TOF camera \emph{can sense} to be rigid. For many tasks that include simultaneous localization and mapping (SLAM), obstacle detection and avoidance, and object manipulation, this is a reasonable assumption \cite{Henry2010,Rangel-Butanda2012,Fernandez-Madrigal2013}.

To ensure that the overall power for depth sensing is actually reduced, we account for the computation power and require our algorithm to estimate accurate depth maps on a low power embedded platform with minimal latency. This means that we cannot blindly use standard techniques to estimate a new depth map because these platforms have limited compute resources. Our contribution therefore is an optimized algorithm that combines computationally efficient techniques to obtain an accurate and dense depth map with minimal latency. Our approach balances the usage the TOF camera, the computational costs of the algorithm, and the quality of the estimated depth. In particular, we present the following: 
\begin{itemize}
	\item We introduce an algorithm that lowers the usage of the TOF camera and instead obtains depth maps by estimating the 3D motion in the scene, which is used to update a previously measured depth map. 
	\item We reduce the computation required to estimate the 3D motion of every pixel by estimating the pose between frames. We show that it is possible to obtain an accurate depth map by using the pose estimated with the optical flow determined by a block matching heuristic on a sparse, uniformly-spaced grid.  This is essential for our approach to run in real time on an embedded platform, which ensures that the overall power for depth sensing is reduced. 
	\item We develop a mechanism to detect when an accurate depth map cannot be estimated and to adaptively enable the TOF camera in these cases. This is crucial because it is not always possible to estimate accurate optical flow (especially with limited compute resources), which is required for our technique. 
\end{itemize}
To demonstrate the effectiveness of our approach and quantify the reduction in power, we implement our algorithm on the ODROID XU-3 board \cite{HardKernel} using only the Cortex-A7 CPUs, which estimates depth maps in real time. In addition to estimating depth maps temporally, we also show how our algorithm can be used to infill depth spatially, making it possible to extend the range of a TOF camera and overcome saturation.

This paper is organized as follows. In Section \ref{sec:related_works}, we describe other related approaches that use images to aid in the estimation of depth maps. This is followed by a presentation of our work, where we first describe how we use the optical flow to estimate the 3D motion in a scene (Section \ref{sec:relative_pose}), and then how we use this to robustly estimate depth maps (Section \ref{sec:methods}). In Section \ref{sec:results}, we evaluate our algorithm on a variety of RGB-D datasets, where we also analyze the tradeoffs of our approach and compare it to other techniques. To estimate the reduction in the power for depth sensing, we quantify the overall system power of our approach in Section \ref{sec:energy}.  In Section \ref{sec:extend}, we show how our algorithm can also be used to infill depth spatially. Finally, we conclude this paper in Section \ref{sec:conclusion}.

\section{Background}
\label{sec:related_works}

The idea of using images to enhance and estimate depth maps has been explored in many applications. For example, the authors of \cite{Lu2011,Park2011,Ferstl,Seung-WonJung2014,Richter2016,Liu2017} use images to upsample low resolution depth maps, and the authors of \cite{JiejieZhu2011,Schwarz2014} enhance TOF camera depth maps using stereo images. Given this breadth, we focus only on techniques that have similar problem setups, namely those that estimate new depth maps temporally using concurrently collected images and previously measured depth maps and those that only use consecutive and monocular images to estimate depth.

\subsection{Temporal Depth Map Estimation}
\label{sec:temp_depth_est}
Here, we survey techniques that use images to temporally estimate new depth maps from previously measured ones. The authors of \cite{Choi2010} address the fact that depth maps obtained from TOF cameras often have lower resolutions and are acquired at lower frame rates than that of digital cameras. To overcome these limitations, they proposed using joint bilateral upsampling techniques to first increase the resolution of the captured depth maps. To estimate the remaining depth maps and equalize the frame rate between the TOF and the digital camera, the authors applied bidirectional block matching algorithms to estimate the optical flow between images without any corresponding depth maps and those with it. These optical flow vectors are used to identify the depth blocks that are averaged to form a new depth map. 

Similar to \cite{Choi2010}, the authors of \cite{Wang2010} and \cite{Zhang2014} also estimate depth maps between frames that have both images and depth maps available using block matching algorithms. However, the authors of \cite{Wang2010} selects the depth block from either the preceding or future depth map based on the edges of the corresponding image blocks. The authors of \cite{Zhang2014} estimate depth by performing a weighted average guided by the underlying texture in the images. 

All of these approaches use block matching algorithms to obtain dense optical flow fields, but this process is computationally expensive. To reduce the complexity, the authors of \cite{Li2011} reuse the motion vectors generated in compressed video to accelerate the process of depth map estimation. In this work, the authors assume that depth maps are acquired only at I- or P-frames and estimate the depth map for B-frames using temporal averaging similar to \cite{Choi2010}.    

Unfortunately, we cannot directly use these techniques to obtain new depth maps because computing a dense optical flow field is \emph{prohibitively slow}. For example, OpenCV's implementation of dense optical flow \cite{Farneback2003} runs at 0.88 frames per second on our embedded device.  For applications like 3D video frame upsampling, which can be performed offline, this is not necessarily a problem, but for applications like robotic navigation, these approaches are unsuitable because the underlying applications are sensitive to latency. Furthermore, most of these approaches also estimate depth maps by using the depth from preceding and future frames. This is not possible for real time applications, and we require the estimation of depth maps to be causal. As we show in Sections \ref{sec:relative_pose} and \ref{sec:methods}, our algorithm uses the assumption of rigidity to significantly reduce the computation required to obtain causal and dense depth maps.

\subsection{Pose Estimation and Structure-from-Motion}
\label{sec:sfm}
As stated in Section \ref{sec:intro}, we use the optical flow to estimate the 3D motion in the scene from frame to frame. For rigid objects, this is represented by the relative pose, which is composed of a rotation and translation. A common way to estimate the pose exploits epipolar geometry and uses the pixel wise correspondences between consecutive images (which can be trivially obtained using the optical flow) to obtain an intermediate quantity known as the essential matrix, which can then factored to obtain the rotation and translation \cite{Hartley2004}. Depending on the number of correspondences, the essential matrix can be estimated using techniques that range from performing a singular value decomposition (8 correspondences) \cite{citeulike:752820} to finding the roots of a tenth order polynomial (5 correspondences) \cite{Nist}. 

One potential benefit of this approach is that it only requires images to obtain pose, although the estimated translation is known only to scale (the magnitude of the translation vector is not known). Furthermore, once the pose is obtained, relative depth can also be estimated by triangulating the corresponding pixels. These techniques are known as structure-from-motion (SfM) \cite{Snavely2006,Wu2013,engel14eccv,schoenberger2016sfm}, and we refer the interested reader to a comparison \cite{Bianco2018} of popular and state-of-the-art pipelines. Unfortunately, one drawback of these approaches is that they typically only estimate depth at a sparse set of keypoints. This is problematic for applications like obstacle avoidance, which require dense depth maps. Furthermore, these techniques also only estimate relative depth.      

Unlike the SfM techniques, our approach uses the relative pose to update a previously measured depth map to obtain a new and \emph{absolute} depth map. We show that using previous depth measurements, which is freely available in our problem setup, allows us to estimate the rotation and {absolute} translation with fewer correspondences, which is important for obtaining accurate depth maps. Previously, we exploited our problem setup to directly estimate the angular and absolute translational velocity, another way of representing rigid motion, using linear least squares \cite{Noraky2017}. Here, we extend our previous approach to estimate the rotation and translation, which allows us to further reduce the usage of the TOF camera with a negligible increase in computation.
 
\section{Relative Pose Estimation}
\label{sec:relative_pose}
In this section, we describe how we estimate the relative pose of a rigid object using the 2D motion of its pixels, or the optical flow, by inverting a simple image formation model. We take a different approach from \cite{Noraky2017} and estimate the rotation and translation. This allows us to further reduce the usage of the TOF camera compared to \cite{Noraky2017} with a negligible increase in complexity. Once the relative pose is obtained, we can determine the 3D motion in the scene and estimate a new depth map, which we describe in Section \ref{sec:methods}. 

Our approach assumes that images are formed by perspective projection. This means that the $i^{\text{th}}$ pixel located at $(u_i,v_i)$ corresponds to the 3D point, $X_i$, in the camera-centric coordinate system:
\begin{equation}\label{eq:perspective2}
	X_i = \frac{z_i}{f}(u_i,v_i,f)^T
\end{equation}
where we denote $z_i$ as the depth of the $i^{\text{th}}$ pixel and $f$ as the focal length. We simplify notation and assume that all image coordinates are relative to the principal point. Given this, we can obtain the 3D position for each pixel in the depth map.

As the object undergoes rigid motion, its motion can be represented by its relative pose, which is composed of a rotation, $R$, and a translation, $T$.  This new 3D point corresponds to the pixel located at $({u}_j,{v}_j)$, where:
\begin{equation}\label{eq:new_pix_coord}
	u_j = f\frac{\hat{x} \cdot (RX_i+T)}{\hat{z} \cdot  (RX_i+T)} \quad \text{and} \quad v_j = f\frac{\hat{y} \cdot  (RX_i+T)}{\hat{z} \cdot  (RX_i+T)}
\end{equation}
Here, we denote $\cdot$ as the dot product and $(\hat{x},\hat{y},\hat{z})$ as the unit vectors oriented along the coordinate axes. 

Given the pixel-wise correspondences between frames, we can obtain the pose by rearranging \eqref{eq:new_pix_coord} and solve for $R$ and $T$ in a least squares sense. Because rotation matrices are nonlinear, we must solve for the pose iteratively. However, instead of estimating the rotation matrix, we use a more compact representation for rotation, namely Rodrigues' Formula \cite{rovis}, where:
\begin{equation}\label{eq:rodrigues}
	R = I + {\sin\theta K+(1-\cos\theta)K^2}
\end{equation}
This describes a rotation of $\theta$ radians about an axis, $\hat{k}$. The vector $\hat{k}$ is a unit vector, whose elements form the skew-symmetric matrix, $K$, such that $KX_i = \hat{k}\times X_i$ (where $\times$ denotes the cross product), and $I$ is the identity matrix. 

We substitute \eqref{eq:rodrigues} into \eqref{eq:new_pix_coord} and rearrange the terms to obtain the following expression that relates the pixel-wise motion, denoted as $\Delta u_i = u_j-u_i$ and $\Delta v_i=v_j-v_i$, to the pose:
\begin{equation}\label{eq:mf_u}
	\Delta u_i =  {\frac{f}{z_i}\hat{x}\cdot\left(WX_i+T\right)-\frac{u_j}{z_i}\hat{z}\cdot\left(WX_i+T\right)}
\end{equation}
\begin{equation}\label{eq:mf_v}
	\Delta v_i  = {\frac{f}{z_i}\hat{y}\cdot\left(WX_i+T\right)-\frac{v_j}{z_i}\hat{z}\cdot\left(WX_i+T\right)}
\end{equation}
where $W = \sin\theta K+(1-\cos\theta)K^2$. We can then solve for the pose ($\hat{k}$, $\theta$ and $T$) in a least squares sense by minimizing the mean residual ($r_i$) error over the $N$ optical flow estimates:
\begin{equation}\label{eq:obj}
	\min_{\hat{k},\theta,T}  \frac{1}{N}\sum_{i=1}^N \underbrace{\left(\Delta u_i - \Delta\hat{ u}_i\right)^2 + \left(\Delta v_i - \Delta\hat{  v}_i\right)^2}_{r_i}
\end{equation}   
where we denote $\Delta\hat{ u}_i$ and $\Delta\hat{ v}_i$ as the right hand side of \eqref{eq:mf_u} and \eqref{eq:mf_v}, respectively. 

We minimize \eqref{eq:obj} using a variant of the Gauss-Newton algorithm, where we linearize the non-linear residual using the Jacobian at $\theta=0$. This assumes that the rotation between frames is small, which is a reasonable assumption for many indoor applications, where images are acquired at 30 frames per second. Furthermore, each iteration of the Gauss-Newton algorithm is equivalent to the least squares solution presented in \cite{Noraky2017}, which is computationally simple and equivalent to solving a $6\times6$ linear system. To estimate the pose, we need at least 3 optical flow vectors and its corresponding depth. This is advantageous because it reduces the computation required to obtain a depth map compared to some of the methods in Section \ref{sec:related_works} that require dense optical flow.

\section{Proposed Algorithm}

\label{sec:methods}
\begin{figure}[h]
	\centering
	\includegraphics[scale=0.4]{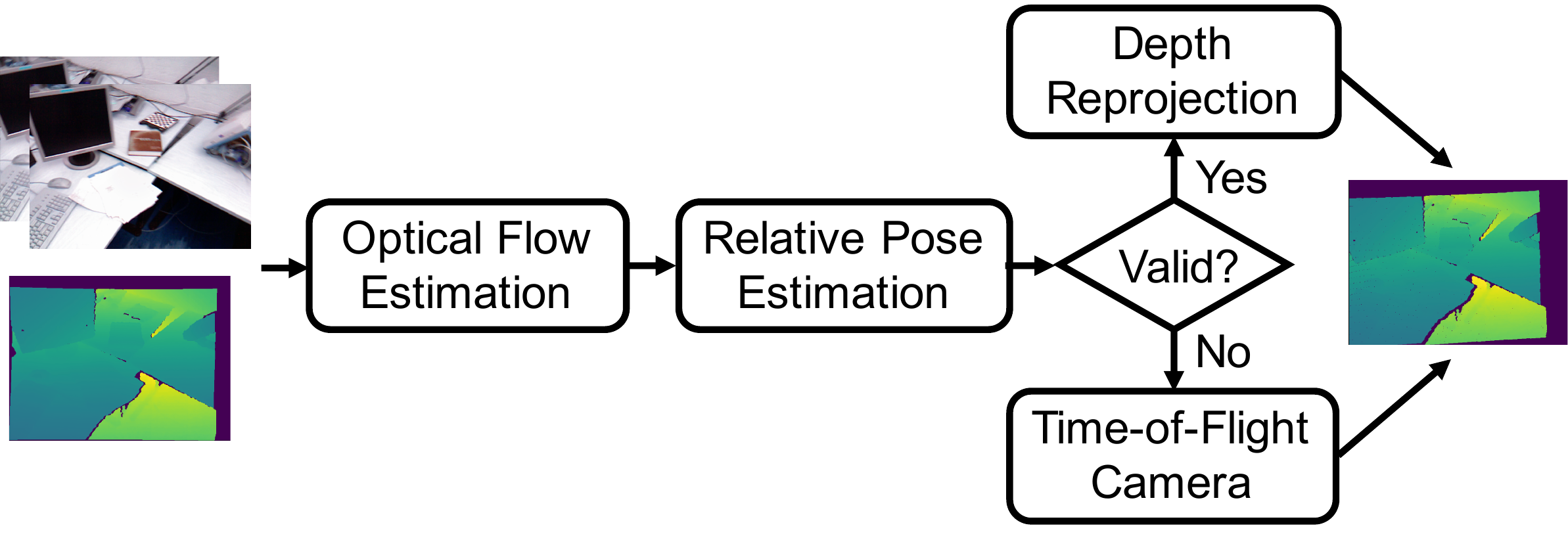}
	\caption{\textbf{Depth Map Estimation Pipeline}: Our algorithm estimates a new depth map using consecutive images and a previously measured depth map. When a reliable depth map cannot be estimated, we use the TOF camera to obtain depth. }
	\label{fig:pipeline}
\end{figure}

In this section, we describe our proposed algorithm, which takes as input consecutive images and a previous depth map and outputs a new one as shown in Figure \ref{fig:pipeline}. Our proposed technique is computationally efficient and we highlight our design choices so that our algorithm can run in real time on an embedded platform. We also describe our strategy to adaptively use the TOF camera when an accurate depth map cannot be estimated.    

\subsection{Optical Flow Estimation}

As shown in Figure \ref{fig:pipeline}, we begin by first estimating the optical flow between consecutive images using the three step search (TSS) algorithm \cite{tss}. The TSS algorithm obtains the optical flow for a block of pixels in an image by searching for the block in the next image that minimizes a cost function.  However, instead of an exhaustive search, the TSS algorithm only considers select locations to reduce computation. In our implementation, we search for the block that minimizes the sum of absolute differences using $15\times15$ blocks with a step size of 8. We also only compute the optical flow for the pixels on a $12\times12$ grid that is uniformly spaced across the image. This reduces the computation because our technique does not require dense optical flow estimates, and we do not need to find keypoints or corners.  

Our decision to use the TSS algorithm is motivated by its run time on an embedded platform. We compare the run time of the TSS algorithm to the commonly used Lucas Kanade algorithm \cite{lucaskanade} by profiling both approaches on the ODROID-XU3 board \cite{HardKernel}, which is an embedded platform that is representative of the compute resources available on mobile devices. We use the Cortex-A7 cores to compute the optical flow for $640\times480$ images for the pixels on a uniformly spaced $12\times12$ grid. For the Lucas Kanade algorithm, we used $15\times15$ blocks and 3 pyramid levels. 

On average, we find that the TSS algorithm require {13 ms} whereas the Lucas Kanade algorithm requires {51 ms}. We also profile the time required to identify corners. We found that the Harris corner detector \cite{HarrisCorner} requires 120 ms, which is intolerable for real time applications, whereas the time to locate the pixels on a uniform grid is negligible. We summarize these run times in Table \ref{tab:runtime}.

\begin{table}[h]
	\centering
	\begin{tabular}{lr}
	\toprule
	\textbf{Algorithm} & \textbf{Runtime (ms)}  \\
	\midrule
	Three Step Search & 13  \\
	Lucas Kanade & 51  \\
	Harris Corner & 120 \\
	\bottomrule
	\end{tabular}
	\caption{\textbf{Runtime Comparisons}: We profile our design choices on the ODROID-XU3 board \cite{HardKernel}. We opt to use the TSS algorithm to ensure our implementation can estimate depth maps in real time.}\label{tab:runtime}
\end{table}

\begin{figure}
	\centering
	\begin{subfigure}{.25\textwidth}
  		\centering
  		\includegraphics[scale=0.75]{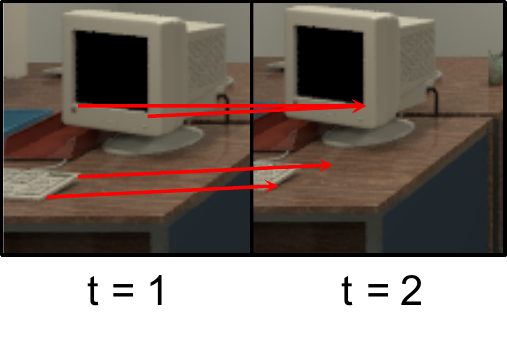}
  		\caption{TSS}\label{fig:tss_output}
	\end{subfigure}%
	\begin{subfigure}{.25\textwidth}
  	\centering
  		\includegraphics[scale=0.75]{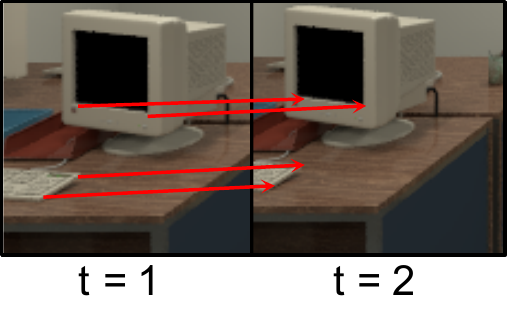}
  		\caption{Using Pose}\label{fig:thiswork_output}
	\end{subfigure}
	\caption{\textbf{Select Optical Flow}: We show examples of optical flow vectors estimated using the TSS algorithm and those obtained using the estimated pose.}\label{fig:test}
\end{figure}

However, as shown in Figure \ref{fig:tss_output}, one drawback of using the TSS algorithm is that our optical flow estimates can be inaccurate. In the next section, we show how we can mitigate this to robustly estimate the pose. With the pose estimated, in addition to obtaining a new depth map, we can also correct the optical flow field as shown in Figure \ref{fig:thiswork_output}.

\subsection{Relative Pose Estimation}
\label{sec:relpose}
With the optical flow estimated, we can solve for the pose as described in Section \ref{sec:relative_pose}. However, using the optical flow directly is problematic because it can be different from the underlying motion field \cite{rovis}. Image sensor noise, occlusions, and the algorithm used to estimate the optical flow affect the accuracy of the estimates. Furthermore, because optical flow is estimated using image intensities, it can be different from the underlying motion field even in the absence of these issues. In regions with uniform intensity, for example, the optical flow would be zero even when the underlying motion field is not. Moreover, the depth values can also be affected by sensor noise in addition irregularities that arise from multipath reflections,  specular reflections, and interference \cite{Hansard2013}.  Because our pose estimation directly uses the depth, these errors can also adversely affect the depth map. 

While these errors are in part mitigated by our least squares formulation, we need a mechanism to distinguish accurate optical flow and depth from erroneous ones because the squared penalty in our formulation is not robust against outliers. This is possible when the pose is known, and we can distinguish the accurate optical flow estimates, or inliers, from the erroneous outliers because the former satisfy \eqref{eq:mf_u} and \eqref{eq:mf_v}. This insight suggests that we solve \eqref{eq:obj} using RANSAC \cite{Fischler1981} to iteratively estimate both the pose and the set of inliers. 

We proceed by randomly selecting the optical flow estimates and its corresponding depth to obtain an initial pose hypothesis. We use 3 optical flow estimates, which is the minimum required to estimate pose using our technique, to minimize the likelihood of choosing an outlier. To judge the quality of the pose hypotheses, we relax the requirement that the pose must satisfy \eqref{eq:mf_u} and \eqref{eq:mf_v} for all of the inliers and instead compute the residual error for each optical flow estimate. If the number of optical flow estimates with low residual errors, which is determined by a threshold, exceed a fraction of the total number of estimates, we resolve \eqref{eq:obj} using only these inliers. We repeat this procedure and select the candidate pose with the lowest mean residual error. When there are no candidates, we enable the TOF camera to acquire a new depth map. This adaptive control of the TOF camera is different from what we presented in \cite{Noraky2017} and allows for robust depth map estimation. We summarize our approach in Algorithm \ref{alg:pose_est}.

To show that Algorithm \ref{alg:pose_est} can mitigate the impact of errors in depth and the optical flow, we first simulate the idealized depth and optical flow for a given pose and corrupt a subset of them. To reflect the fact that our approach uses the TSS algorithm, we also round each optical flow vector to the nearest integer displacement. We then estimate the pose with and without RANSAC and compare it to the pose we used to simulate the data with using the root mean squared error (RMSE) of the translation as defined in \cite{Sturm2012}.  In Table \ref{tab:ransac_noise}, we see that RANSAC substantially lowers the RMSE of the translation across all scenarios. 

\begin{table}
	\centering
	\begin{tabular}{ccc}
	\toprule
	\textbf{Depth} & \textbf{Optical Flow}& \textbf{Reduction (\%)}  \\
	\midrule
	$\times$ & & 68.0  \\
	& $\times$ & 59.4 \\
	$\times$ & $\times$ & 45.1 \\
	\bottomrule
	\end{tabular}
	\caption{\textbf{Impact of RANSAC}: We present the reduction in the RMSE obtained using RANSAC when there is noise in the depth measurements, the optical flow, and in both.}\label{tab:ransac_noise}
\end{table}

In our implementation, we use 30 RANSAC iterations and set the threshold to 4 and accept a pose hypothesis if the size of its inlier set is at least 10\% of the number of optical flow estimates. When obtaining the initial pose, we only perform 1 iteration of the Gauss-Newton algorithm, which is equivalent to the method presented in \cite{Noraky2017}. We resolve \eqref{eq:obj} using the inlier set by performing 3 iterations. This is negligibly more computation than \cite{Noraky2017}, and as we describe in Section \ref{sec:implementaiton}, the pose estimation accounts for a small fraction of the run-time. This is significant because it allows us to lower of the complexity of the optical flow estimation algorithm and still obtain accurate pose estimates. This is essential to obtaining accurate depth maps in real time. 

\begin{algorithm}
\caption{Adaptive Pose Estimation}
\label{alg:pose_est}
	\begin{algorithmic}[1]
		\Require Optical flow $(\Delta u_i, \Delta v_i)$, depth ($z_i$), and RANSAC parameters (No. of iterations, thresh, and min. size)
		\Ensure Pose ($R$ and $T$) or signal to use TOF camera
		\newline
		\Repeat\Comment{Get the inlier set}
			\State Randomly sample 3 optical flow vectors and its depth
			\State Solve \eqref{eq:obj}; Compute residuals, $r_i$
			\State Get inlier set, $\mathcal{I} = \{i : r_i < \text{thresh}\}$ 
			\State Retain $\mathcal{I}$ with lowest mean residual; $|\mathcal{I}| > $ min. size 
		\Until{End of RANSAC} 
		\newline
		\If{$|\mathcal{I}| = {0} $ }\Comment{Get pose or depth map}
			\State Use the TOF camera
		\Else
			\State Solve \eqref{eq:obj} using $\mathcal{I}$
		\EndIf
	\end{algorithmic}
\end{algorithm}

\subsection{Depth Reprojection}
\label{sec:reproject_sec}

Once the pose is estimated, we obtain a new depth map by applying the pose to each 3D point in the first depth map and projecting its depth, or its $z-$coordinate, to an image. For every pixel in the first depth map, we first compute its 3D point, $X_i$, using \eqref{eq:perspective2}. The reprojected depth map is then obtained as follows: 
\begin{equation}\label{eq:reproject}
	D\left[f\frac{\hat{x} \cdot (RX_i+T)}{\hat{z} \cdot  (RX_i+T)} ,f\frac{\hat{y} \cdot  (RX_i+T)}{\hat{z} \cdot  (RX_i+T)}\right] = \hat{z} \cdot (RX_i+T)
\end{equation}
where $D$ represents the depth map whose entries are indexed by its $x$ and $y$ coordinates. If multiple points are mapped onto the same pixel location, we retain the smallest depth value. 

When more than one depth map is predicted consecutively, we obtain a new depth map by reprojecting the last measured depth map. To do so, we update the pose accordingly. Let $R_c$ and $T_c$ represent the current pose that is estimated using the previously estimated depth map. We also assume that the previously estimated depth map was obtained by reprojecting the last measured depth map using $R_{t-1}$ and $T_{t-1}$. Then, the pose which we now use to reproject the previously {measured} depth map, denoted as $R_t$ and $T_t$, is:
\begin{equation}\label{eq:reproject2}
	R_{t} = R_cR_{t-1} \qquad T_{t} = T_c+R_cT_{t-1}
\end{equation}

\begin{figure}
	\centering
	\includegraphics[scale=0.6]{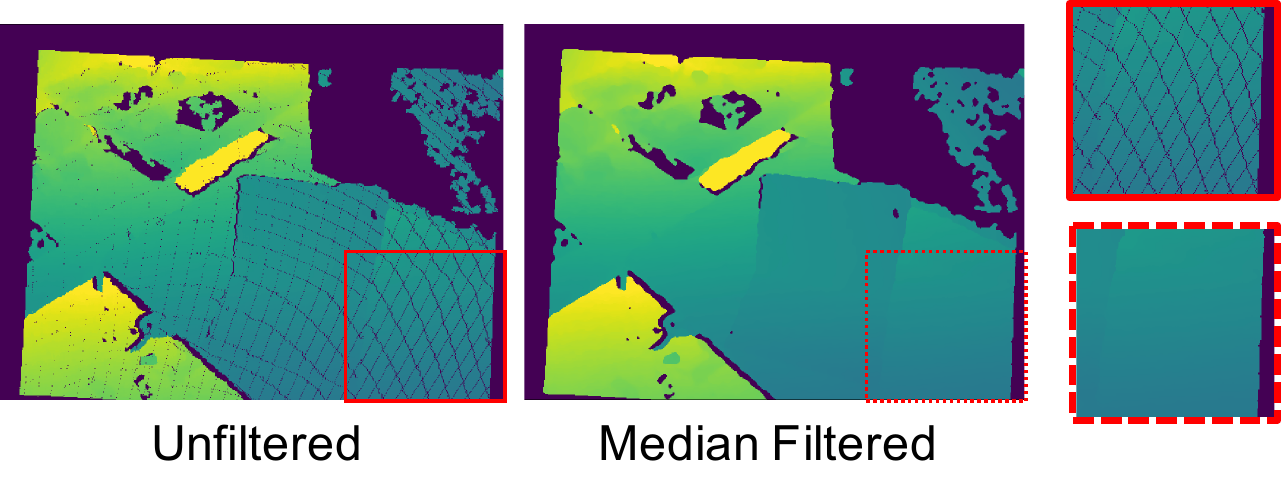}
	\caption{\textbf{Reprojected Depth Maps}: The reprojected depth maps have artifacts where depth is not available. While median filter can infill these regions, we ignore this post-processing step because the holes constitute a small portion of the depth map.}
	\label{fig:holes}
\end{figure}

The resulting depth map contains depth estimates for pixels that correspond to the overlapping field of views between the image where the last depth map was measured and the current image. It should be noted that without any additional post-processing, this method also introduces artifacts as shown in Figure \ref{fig:holes}. These holes arise because the pixels belonging to the same object are treated independently and are not constrained to be contiguous after reprojection and because regions that were previously occluded have been uncovered. While reverse warping would eliminate these holes, it also erroneously infills the previously occluded regions. We want to avoid this, especially as we predict many depth maps consecutively, in the event where the previously occluded region has a different depth from its surroundings. We confirm this by applying our algorithm to sequences from the TU Munich RGB-D dataset \cite{Sturm2012} and find that reverse warping increases the overall mean relative error (as defined in Section \ref{sec:metrics}) by 17.4\% compared to our approach.  Furthermore, if the application needs the depth in the previously occluded regions, this could serve as another signal to use the TOF camera. 

One potential way to remove the first type of holes is by applying a median filter with a small kernel size to the resulting depth map as shown in Figure \ref{fig:holes}. While this may give inconsistent behaviors at depth boundaries, we find in our experiments with the TU Munich RGB-D dataset, the overall mean relative error remains unchanged. However, as our computational resources are limited, we ignore this additional step because these types of holes are minimal, accounting for less than 3\% of the estimated pixels while imposing an additional 20 ms overhead. In the next section, we see that this is intolerable for real-time performance.  

\section{Algorithm Evaluation}
\label{sec:results}

\subsection{Implementation}
\label{sec:implementaiton}
We implement our algorithm on the ODROID XU-3 board \cite{HardKernel}, which is an embedded platform with an Exynos 5422 processor. The Exynos processor is used in the Samsung Galaxy S5 \cite{Samsung} and is representative of the compute power available on mobile devices. Our implementation uses the Cortex-A7 cores of the board and outputs $640\times480$ depth maps in real time, or 30 frames per second (FPS). To achieve this frame rate, we parallelize our computation across the 4 Cortex-A7 cores. We use the parameter settings described in Section \ref{sec:methods} and the OpenCV library whenever possible.

\begin{figure}
	\centering
	\includegraphics[scale=0.3]{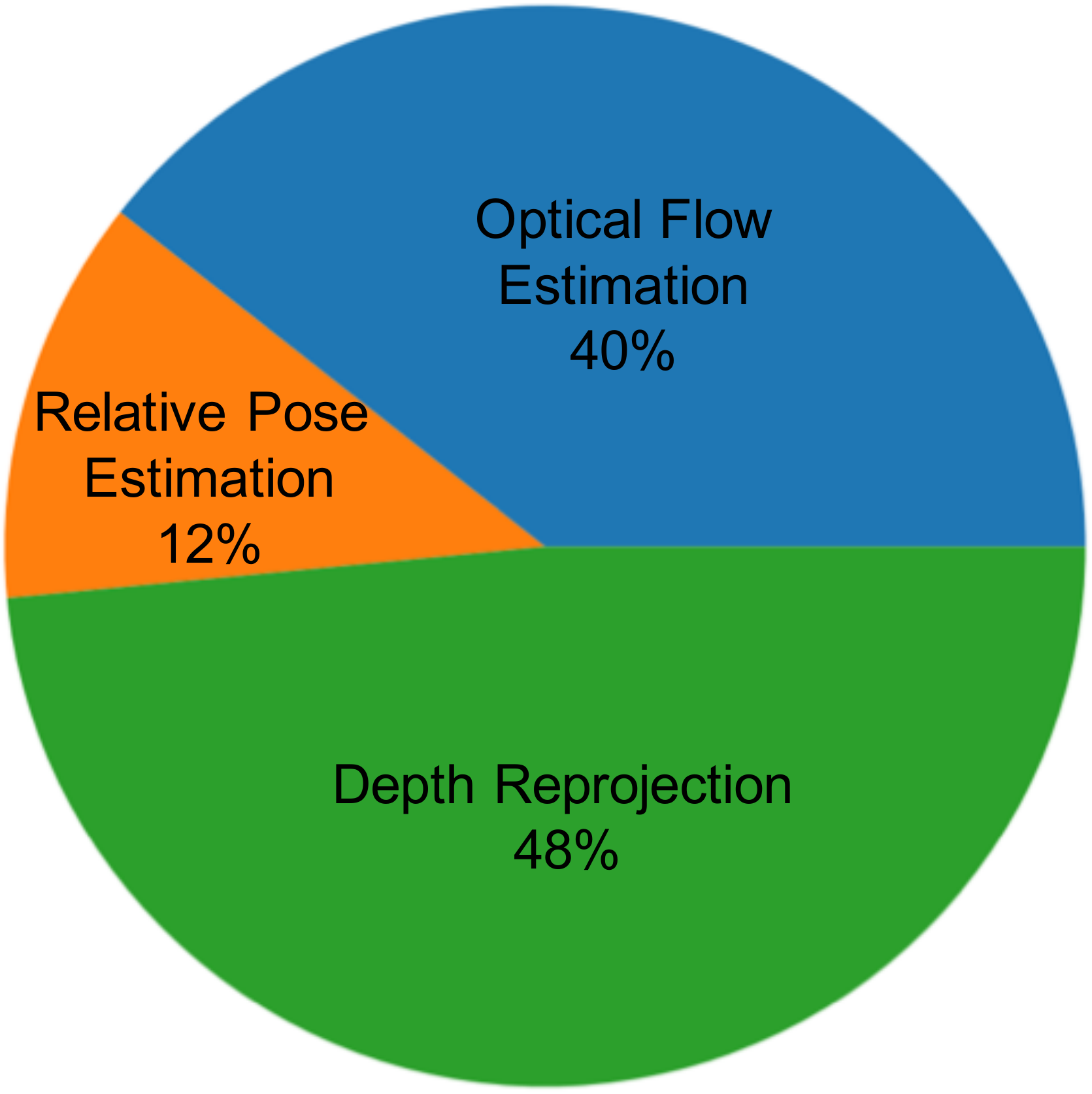}
	\caption{\textbf{Runtime Breakdown}: We profile the implementation of our algorithm on the ODROID-XU3 \cite{HardKernel} board, which produces $640\times480$ depth maps at 30 FPS. Because the time to reproject a new depth map is fixed, we work to reduce the computation time required to estimate the optical flow.}
	\label{fig:time_breakdown}
\end{figure}

As shown in Figure \ref{fig:time_breakdown}, most of the time of our implementation is spent on estimating the optical flow and reprojecting the depth map. This figure further justifies our decision to use the TSS algorithm. Since the time required to reproject a depth map is fixed, we are limited in what we can allocate to obtain the optical flow if we want to estimate depth maps at 30 FPS. We discuss the impact of this decision on the accuracy of the estimated depth maps in Section \ref{sec:discussion1}. This figure also shows that when only the pose is required, which is the case for SLAM, our algorithm can run at nearly 58 FPS.  

\subsection{Dataset}
\label{sec:datasets}
We evaluate our algorithm on RGB-D datasets used to benchmark SLAM, visual odometry, 3D reconstruction, and navigation algorithms. These tasks are relevant for many mobile applications, and the images and depth maps are representative of what our approach will encounter. We adapt these datasets to test our approach by using consecutive images and select depth maps to predict new ones, which we then compare to that in the dataset. For our experiments, we use the provided intrinsic parameters and tools to synchronize the images with the depth maps for each dataset.

We use  sequences from the following datasets: TU Munich RGB-D \cite{Sturm2012}, NYU Depth V2 \cite{Silberman:ECCV12}, Indoor RGB-D \cite{Schmidt2013}, CoRBS \cite{wasenmueller2016corbs}, and ICL-NUIM \cite{handa:etal:ICRA2014}. These datasets contain $640\times480$ RGB images and depth maps and most are collected at 30 FPS.

\subsection{Methodology}
\label{sec:metrics}
We apply our algorithm to the first 100 frames of the sequences in each dataset. We quantify the accuracy of the depth maps using the following error metrics:
\begin{itemize}
	\item \textbf{Mean Relative Error (MRE)}: This is defined as $\frac{100}{N}\sum_{j=1}^N \frac{|z_j-\hat{z}_j|}{z_j}$, where $N$ is the number of pixels predicted, $\hat{z}_j$ is the predicted depth for the $j^{\text{th}}$ pixel, and $z_j$ is the depth measured by the TOF camera. The MRE is presented as a percentage.
	\item \textbf{Mean Absolute Error (MAE)}: This is defined as $\frac{1}{N}\sum_{j=1}^N {|z_j-\hat{z}_j|}$ and presented in centimeters.
	\item \textbf{Root Mean Squared Error (RMSE)}:  This is defined as $\sqrt{\frac{1}{N}\sum_{j=1}^N {(z_j-\hat{z}_j)^2}}$ and presented in centimeters.
\end{itemize}
Because our algorithm uses the TOF camera adaptively, we also quantify the frequency at which it is used using:
\begin{itemize}
	\item \textbf{Duty Cycle (DC)}: This is equal to $\sum_{i=1}^{100}\mathbbm{1}(i)$, where $\mathbbm{1}(i)$ equals 1 if the $i^{\text{th}}$ depth map is obtained using the TOF camera and 0 if the depth map is estimated instead. The DC is presented as a percentage.
\end{itemize}
To reduce the power required to obtain accurate depth maps, our goal is to lower the TOF camera's duty cycle while maintaining a baseline accuracy. In our analysis, we focus on the MRE because it allows us to compare the performance of our algorithm across datasets that have different ranges of depth. Therefore, for each dataset, we set the threshold parameter in Algorithm \ref{alg:pose_est} to achieve a median MRE of approximately 1\% across its sequences in order to measure its duty cycle.   
 
 \subsection{Results}
 \begin{figure}[t!]
    \centering
        \includegraphics[scale=0.4]{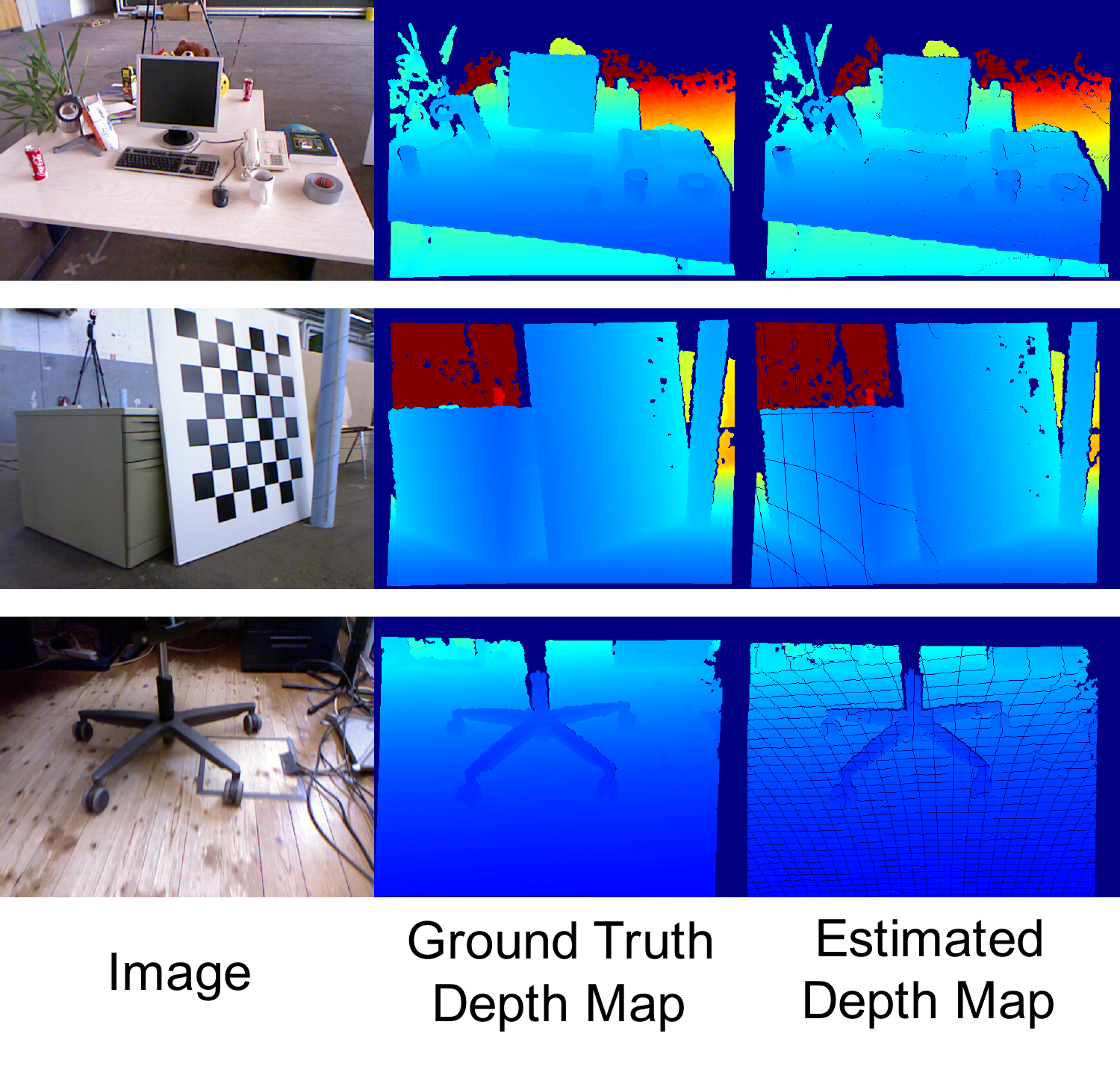}
        \caption{\textbf{Estimated Depth Maps}: We show the estimated depth maps for select sequences in \cite{Sturm2012}. A video of our algorithm running can be found in \cite{youtube}. }
        \label{fig:example}
\end{figure}

We summarize the performance of our algorithm for each dataset in Table \ref{tab:eval}, where we compute the median of each error metric across the depth maps. Examples of the estimated depth maps are shown in Figure \ref{fig:example}. Across the datasets, we achieve a median MRE of 0.96\% and a median duty cycle of 15.0\%. In Table \ref{tab:eval}, we see that the duty cycle for Indoor RGB-D \cite{Schmidt2013} is higher than that of the other datasets. This is expected because this dataset contains sequences of a robot moving abruptly in a sparsely textured environment. Furthermore, this shows that our technique can adapt to and still reduce the usage of the TOF camera in these challenging scenarios. 
\begin{table}
	\centering
	\begin{tabular}{lcccc} \hline
		\toprule
		\textbf{Dataset} & \textbf{MRE (\%)} & \textbf{MAE (cm)} & \textbf{RMSE (cm)} &  \textbf{DC (\%)} \\ 
		\midrule
			TU Munich RGB-D \cite{Sturm2012} & 0.96 & 2.27 & 7.63 & 16.0 \\
			NYU Depth V2  \cite{Silberman:ECCV12} & 0.95 & 4.04 & 9.01& 10.0 \\
			Indoor RGB-D \cite{Schmidt2013} & 1.03 & 2.11 & 7.54 & 33.0 \\
			CoRBS \cite{wasenmueller2016corbs} & 1.04 & 1.79 & 8.98 & 15.0 \\
			ICL-NUIM \cite{handa:etal:ICRA2014} & 0.67 & 2.04 & 5.65 & 10.0 \\
		\midrule
			\textbf{Mean} & 0.93 & 2.45 & 7.76 & 16.8 \\
			\textbf{Median} & 0.96 & 2.11 & 7.63 & 15.0 \\
		\bottomrule		
	\end{tabular}
	\caption{\textbf{Algorithm Evaluation}: We summarize the MRE, MAE, RMSE and DC that our algorithm achieves.}\label{tab:eval}
\end{table}

Because different applications have different accuracy requirements for depth maps, we also quantify the tradeoff between the duty cycle and the MRE for our approach. To do so, we vary the threshold in Algorithm \ref{alg:pose_est} that determines if an optical flow estimate is an inlier. In our pipeline, we expect that a lower threshold, which assumes accurate optical flow estimates, will result in depth maps with a lower MRE but also a higher duty cycle because the TSS algorithm cannot consistently obtain accurate optical flow estimates. By the same reasoning, we expect the MRE to be higher but the duty cycle to be lower when the threshold is high. We present this tradeoff in Figure \ref{fig:tradeoff}, where each point labeled \emph{This Work} in the legend represents the median duty cycle and MRE pair across all of the sequences in each dataset for different thresholds. 

\subsection{Impact of Optical Flow Algorithm}

\label{sec:discussion1}
To quantify the impact of the TSS algorithm on the overall accuracy of estimated depth maps, we compare our algorithm to a variant that uses the Lucas Kanade algorithm to estimate optical flow. We expect the TSS algorithm to perform worse than the Lucas Kanade algorithm in estimating the optical flow because the TSS algorithm only considers select locations in its search for the best matching block, and to increase the overall MRE of the estimated depth map. In Table \ref{tab:alg_comparison}, we compare this variant (\emph{LK}) to our approach (\emph{This Work}) and present the MRE for the same duty cycle.

From this comparison, we see that our hypothesis is confirmed and that using the Lucas Kanade algorithm in our pipeline reduces the overall median MRE from 0.96\% to 0.86\%. However, while the Lucas Kanade algorithm reduces the MRE of the estimated depth maps by over 10\%, it does not justify the 50\% decrease in the estimation frame rate when profiled on the ODROID board. As shown in Table \ref{tab:alg_runtime_comparison}, its frame rate is 15 FPS, which is intolerable for real time applications. 

\subsection{Benefit of Scene Adaptive Estimation}
\label{sec:sadapt}
\begin{figure*}
	\centering
	\begin{subfigure}[b]{0.3\textwidth}\includegraphics[width=\textwidth]{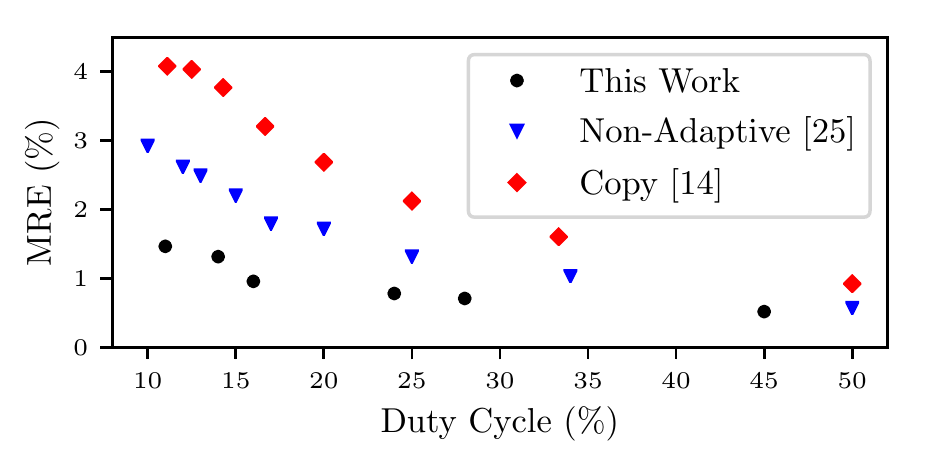}\subcaption{TU Munich RGB-D \cite{Sturm2012}}\end{subfigure}
	\begin{subfigure}[b]{0.3\textwidth}\includegraphics[width=\textwidth]{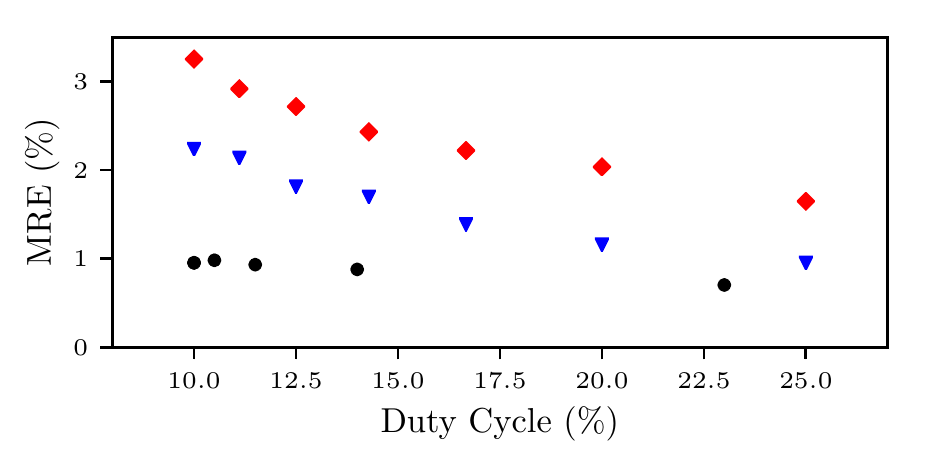}\subcaption{NYU Depth V2  \cite{Silberman:ECCV12}}\end{subfigure}
	\begin{subfigure}[b]{0.3\textwidth}\includegraphics[width=\textwidth]{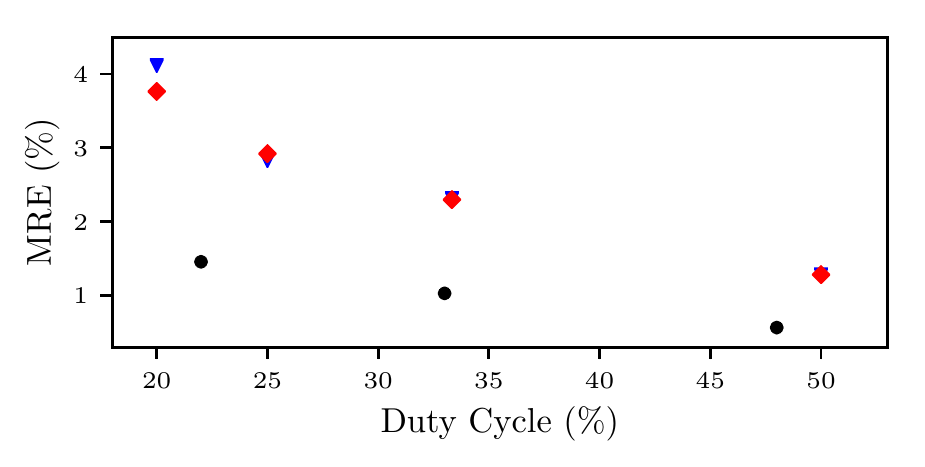}\subcaption{Indoor RGB-D\cite{Schmidt2013}}\end{subfigure}
	\medskip
	\begin{subfigure}[b]{0.3\textwidth}\includegraphics[width=\textwidth]{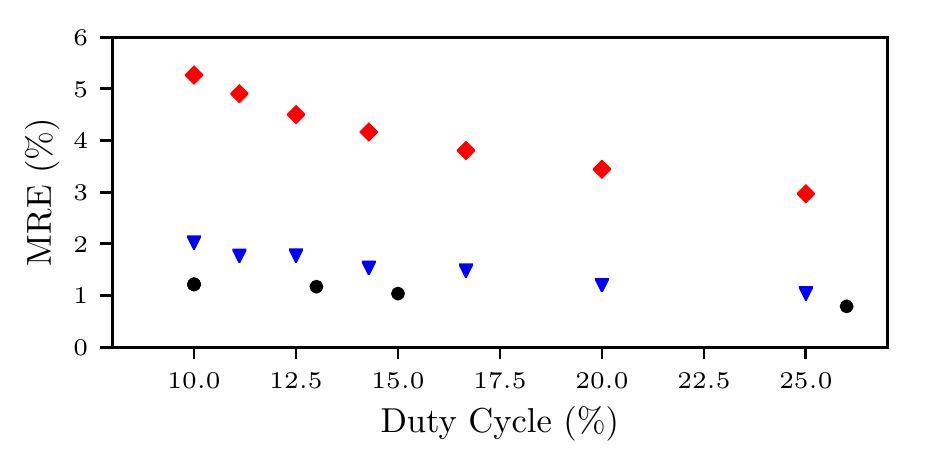}\subcaption{CoRBS \cite{wasenmueller2016corbs}}\end{subfigure}
	\begin{subfigure}[b]{0.3\textwidth}\includegraphics[width=\textwidth]{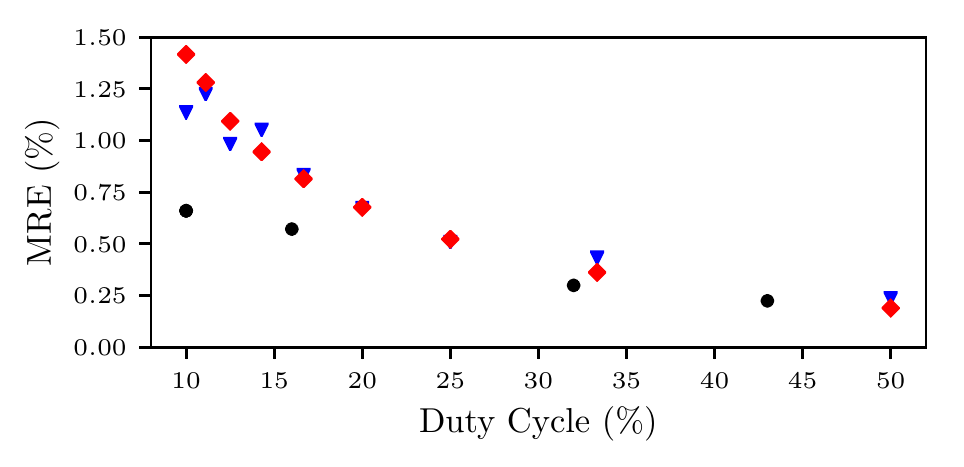}\subcaption{ICL-NUIM \cite{handa:etal:ICRA2014}}\end{subfigure}
	\caption{\textbf{Tradeoff Between Duty Cycle and MRE}: We show the tradeoff between the duty cycle and MRE for our technique (This Work) and the techniques we compare against: Non-Adaptive (Section \ref{sec:sadapt}) and Copy (Section \ref{sec:comparison1}). Because our technique is adaptive, our duty cycles do not align with the competing techniques, which estimate depth at regular intervals. }\label{fig:tradeoff}
\end{figure*}

One key feature of our algorithm is that it adaptively uses the TOF camera when an accurate depth map cannot be estimated. This is necessary because it is not always possible to obtain accurate optical flow estimates. We compare our adaptive scheme to our previous work \cite{Noraky2017}, which predicts depth maps at regular intervals. We apply this approach to the datasets in Section \ref{sec:datasets} and plot the duty cycle and MRE pairs, which are denoted as \emph{Non-Adaptive} in Figure \ref{fig:tradeoff}.

In this figure, we see that the adaptive scheme of our approach outperforms \cite{Noraky2017} across all duty cycles with a negligible increase in complexity. For the same duty cycle, we see in Table \ref{tab:alg_comparison} that our adaptive schemes reduces the median MRE from 1.80\% to 0.96\%. Furthermore, this result make sense upon inspecting the images in the datasets. Images with rapid motion are blurred and contain large displacements, making the estimation of accurate optical flow challenging. Our algorithm is optimized to detect these scenarios and uses the TOF camera while estimating depth maps for frames with slower motion.  

\begin{table*}[h]
	\centering
	\begin{tabular}{lccccccccccccccc} \hline
		\toprule
		\textbf{Dataset} & \multicolumn{3}{c}{\textbf{This Work}} & \multicolumn{3}{c}{\textbf{LK}}  & \multicolumn{3}{c}{\textbf{Non-Adaptive \cite{Noraky2017}}} & \multicolumn{3}{c}{\textbf{Copy \cite{Wang2010}}} & \multicolumn{3}{c}{\textbf{SfM-SIFT \cite{Bianco2018}}} \\ 
		\midrule
		& \tiny MRE (\%) &  \tiny MAE (cm) &  \tiny RMSE (cm)  & \tiny MRE (\%) &  \tiny MAE (cm) &  \tiny RMSE (cm) & \tiny MRE (\%) &  \tiny MAE (cm) &  \tiny RMSE (cm)  & \tiny MRE (\%) &  \tiny MAE (cm) &  \tiny RMSE (cm)  & \tiny MRE (\%) &  \tiny MAE (cm) &  \tiny RMSE (cm)  \\
		\midrule
			TU Munich RGB-D \cite{Sturm2012}& 0.96	&	2.27	&	7.63	&	0.86	&	1.68	&	7.55	&	1.80	&	6.26	&	13.83	&	3.20	&	5.80	&	25.97	&	36.14	&	83.77	&	104.99	\\
			NYU Depth V2  \cite{Silberman:ECCV12} & 0.95	&	4.04	&	9.01	&	0.82	&	3.34	&	8.11	&	2.24	&	5.98	&	14.65	&	3.25	&	10.04	&	40.25	&	43.03	&	171.00	&	212.91	\\
			Indoor RGB-D \cite{Schmidt2013} & 1.03	&	2.11	&	7.54	&	1.49	&	3.53	&	13.20	&	2.32	&	5.31	&	14.96	&	2.30	&	5.41	&	20.04	&	32.29	&	119.32	&	154.14	\\
			CoRBS \cite{wasenmueller2016corbs} & 1.04	&	1.79	&	8.98	&	1.02	&	1.66	&	9.82	&	1.54	&	2.94	&	12.28	&	4.16	&	9.44	&	34.87	&	38.61	&	80.52	&	106.66	\\
			ICL-NUIM \cite{handa:etal:ICRA2014} & 0.67	&	2.04	&	5.65	&	0.14	&	0.39	&	3.02	&	1.14	&	3.21	&	8.26	&	1.42	&	3.98	&	10.62	&	39.76	&	126.38	&	158.74	\\
		\midrule
			\textbf{Mean}&0.93	&	2.45	&	7.76	&	0.87	&	2.12	&	8.34	&	1.81	&	4.74	&	12.80	&	2.87	&	6.93	&	26.35	&	37.97	&	116.20	&	147.49	\\
			\textbf{Median}&0.96	&	2.11	&	7.63	&	0.86	&	1.68	&	8.11	&	1.80	&	5.31	&	13.83	&	3.20	&	5.80	&	25.97	&	38.61	&	119.32	&	154.14	\\
		\bottomrule		
	\end{tabular}
	\caption{\textbf{Algorithm Comparison}: We compare the performance of our algorithm to variants and competing techniques for approximately the same duty cycle to show that our approach estimates accurate depth maps.}\label{tab:alg_comparison}
\end{table*}

\begin{table}[h]
	\centering
	\begin{tabular}{lc}
	\toprule
	\textbf{Algorithm} & \textbf{Frame Rate (FPS)} \\
	\midrule
	This Work &  30 \\
	LK & 15 \\
	Non-Adaptive \cite{Noraky2017} & 30 \\
	Copy \cite{Wang2010} &   0.83 \\
	SfM-SIFT \cite{Bianco2018} &  0.12\\
	SfM-SURF \cite{Bianco2018} &  0.36\\
	SfM-ORB\cite{Bianco2018} &  1.81\\

	\bottomrule
	\end{tabular}
	\caption{\textbf{Algorithm Frame Rate Comparison}: We compare the estimation frame rates our approach and other techniques on the ODROID-XU3 board \cite{HardKernel}. }\label{tab:alg_runtime_comparison}
\end{table}

\subsection{Comparison to Previous Work}

\subsubsection{Temporal Depth Map Estimation}
\label{sec:comparison1}
We compare our algorithm to a causal variant of \cite{Wang2010} as described in Section \ref{sec:temp_depth_est}. This technique estimates depth by copying previous measurements guided by the optical flow. Since our setup requires depth maps to be estimated in real time, we use the optical flow between the current and preceding images to copy the depth from a previous frame. In our experiments, we compute a dense optical flow field using \cite{Farneback2003}. The estimation of dense optical flow is prohibitively slow on our embedded processor, and this technique runs at 0.83 FPS as shown in Table \ref{tab:alg_runtime_comparison}. However, we still perform this experiment to quantify the effectiveness of remapping depth. We expect this approach to perform well when the motion between frames is small.   

We apply this approach to the datasets and plot the duty cycle and MRE pairs, which we denote as \emph{Copy}, in Figure \ref{fig:tradeoff}. From this figure, we see that our approach outperforms \emph{Copy} across all duty cycles and datasets. This result shows that our dataset contains non-trivial changes in depth that cannot be captured by simply remapping the pixels of a previous depth map. Furthermore, this experiment suggests that the changes in depth can be estimated by our technique.

\subsubsection{Structure-from-Motion} 
\label{sec:comparison2}
We also compare our algorithm to a structure-from-motion (SfM) pipeline that estimates relative depth. Even though SfM estimates relative depth at a sparse set of points, these techniques only use images and can be compelling if it can run in real-time on a low-power embedded platform. We implement an incremental SfM pipeline following standard and state of the art approaches described in \cite{Bianco2018}. We use SIFT \cite{Lowe2004} to localize keypoints, match consecutive keypoints using brute force matching, perform geometric validation using the 8 point algorithm, and triangulate using the DLT method \cite{Hartley2004}. We apply the SfM pipeline to our setup and estimate the depth using two consecutive images. Across the different datasets, our SfM pipeline estimates the depth at approximately 210 keypoints. 

As summarized in Table \ref{tab:alg_runtime_comparison}, our implementation (\emph{SfM-SIFT}) runs at 0.12 FPS on the ODROID XU-3 board, where most of the time is spent on computing and matching the keypoints. Due to the low frame rate, we also experimented with using SURF \cite{Bay2008}  (\emph{SfM-SURF}) and ORB \cite{Rublee2011}  (\emph{SfM-ORB}) features instead of SIFT. These variants estimate sparse depth at 0.36 and 1.8 FPS, respectively. While these variants have a higher frame rate than the standard pipeline, they are still far from real time. To quantify the accuracy of the depth estimates obtained using the standard SfM pipeline, we find the scale factor so that the estimated relative depth best matches the ground truth. We summarize the MRE for each dataset in Table \ref{tab:alg_runtime_comparison}, where we also compare it (\emph{SfM-SIFT}) to our approach and other competing techniques. Because our pipeline uses only two images, the high MRE is expected. We can lower the MRE by incorporating more frames and performing bundle adjustment \cite{Hartley2004}, but this would increase latency and further decrease the estimation frame rate. Due to the high MRE and the low frame rate, we see that SfM is impractical for the scenario we consider.

\section{System Power Reduction}
\label{sec:energy}

To lower the power for TOF imaging, our strategy is to lower the duty cycle of the TOF camera and estimate depth maps instead. However, this implies that the power required to estimate a new depth map is less than that of using a TOF camera. Here, we measure the power of an implementation of our algorithm on the ODROID XU-3 board and use it to estimate the overall system power of a system that uses our algorithm alongside the TOF camera to obtain depth.    

The ODROID XU-3 board has 4 Cortex-A7 CPUs and 4 Cortex-A15 CPUs. To keep the computation power low, we only use the Cortex-A7 cores to estimate the depth maps. This leaves the Cortex-A15 cores available for other mobile applications that use depth maps and further underscores that our implementation, which outputs $640\times480$ depth maps in real time, is efficient. The resulting implementation consumes a total of 0.69 W, of which the idle power is {0.19 W}. We summarize the power breakdown of our implementation in Table \ref{tab:spec}.

\begin{table}
	\centering 
	\begin{tabular}{llr}  
		\toprule
		\multicolumn{2}{c}{\textbf{Category}} & \textbf{Power (W)} \\
		\midrule
		Core      	& Active    & 0.63      \\
			&    Idle     & 0.16       \\
		DRAM       & Active     & 0.06      \\
       			& Idle     & 0.03      \\
		Total & Active      & 0.69       \\
			 & Idle      & 0.19       \\
		\bottomrule
	\end{tabular}
	\caption{\textbf{Power Breakdown}: We measure the power of our implementation on the ODROID-XU3 board \cite{HardKernel}.}\label{tab:spec} 
\end{table}

Given the power of our implementation, we now estimate the overall system power of a hybrid system that uses the TOF camera and our algorithm to obtain depth. We define the overall system power, denoted as $P_S$, as follows: 
\begin{equation}\label{eq:power}
	P_S = \frac{ON}{100}\cdot(P_{TOF}+P_{I}) + (1-\frac{ON}{100})\cdot(P_{C} + P_{M})
\end{equation}
where we denote $ON$ as the duty cycle of the TOF camera, $P_{TOF}$ is the power of the TOF camera, $P_{I}$ is the total idle power, $P_{C}$ is the active power of the A7 cores, and $P_{M}$ is the active power of the DRAM. Because we assume that images are routinely collected for other purposes, we ignore its contribution in \eqref{eq:power}. Based on a survey of commercial TOF cameras (with ranges up to 4 meters), we also assume that $P_{TOF}$ ranges from 1 to 5 W \cite{Bamji2018}\cite{Colaco2013}. 

\begin{figure}
	\centering
	\includegraphics[scale=0.8]{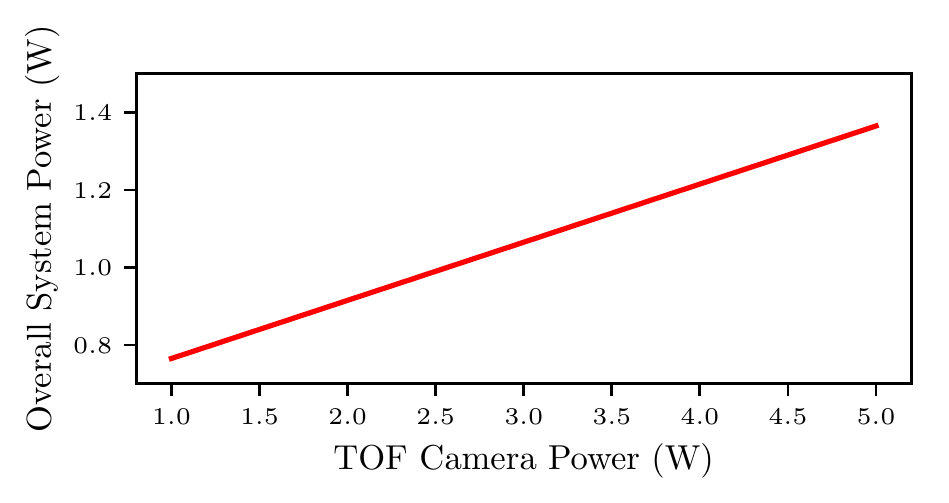}
	\caption{ \textbf{Overall System Power}: We estimate the power of a system that uses our algorithm to estimate depth alongside the TOF camera. For commercial TOF cameras, our algorithm can reduces the overall system power by 23\%-73\%.}
	\label{fig:reduction}
\end{figure}

Taking the duty cycle to be 15\% and using the measurements in Table \ref{tab:spec},  we plot the power of the hybrid system in Figure \ref{fig:reduction}. For the datasets in Table \ref{tab:eval}, this translates to a median power reduction of 23\%-73\% compared to just using the TOF camera while producing depth maps with a median MRE of 0.96\%.

 \section{Infilling Depth Maps}
 \label{sec:extend}
 
In the previous section, we describe how we use our algorithm to estimate new depth maps \emph{temporally} to lower the power for TOF imaging. Here, we show that our algorithm can also be used to estimate depth \emph{spatially} to infill missing depth values. This means that our algorithm can be used to address two deficiencies of TOF imaging, namely when the sensor goes out of range and when the sensor saturates. We consider both cases and show how we can, in effect, extend the range of a TOF camera without increasing the power of its illumination source and overcome saturation.

\begin{figure}
	\centering
	\includegraphics[scale=0.4]{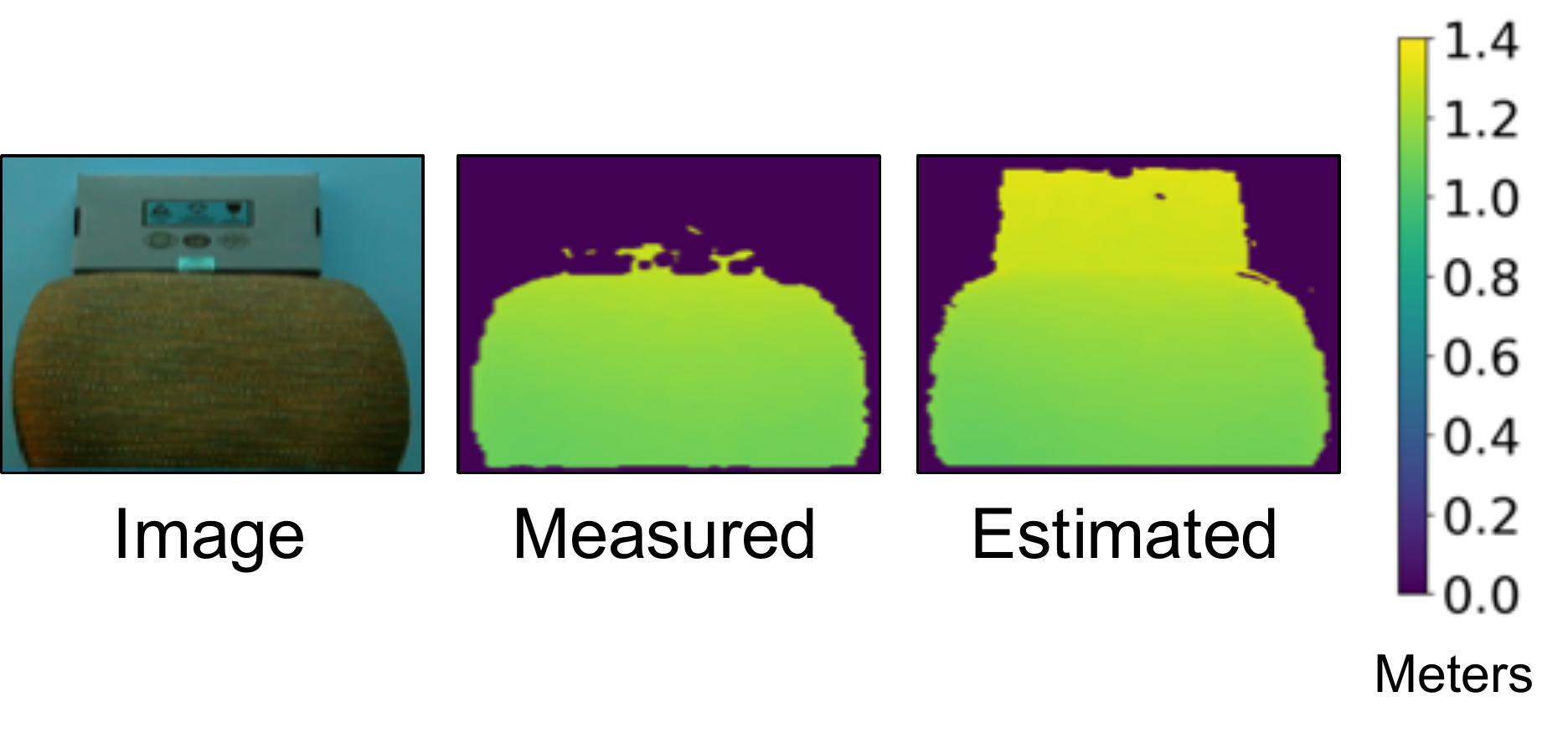}
	\caption{ \textbf{Out of Range}: We estimate the depth for objects that exceed the TOF camera's range using our algorithm. The purple regions cannot be sensed by the TOF camera.}
	\label{fig:outofrange}
\end{figure}

\begin{figure}
	\centering
	\includegraphics[scale=0.4]{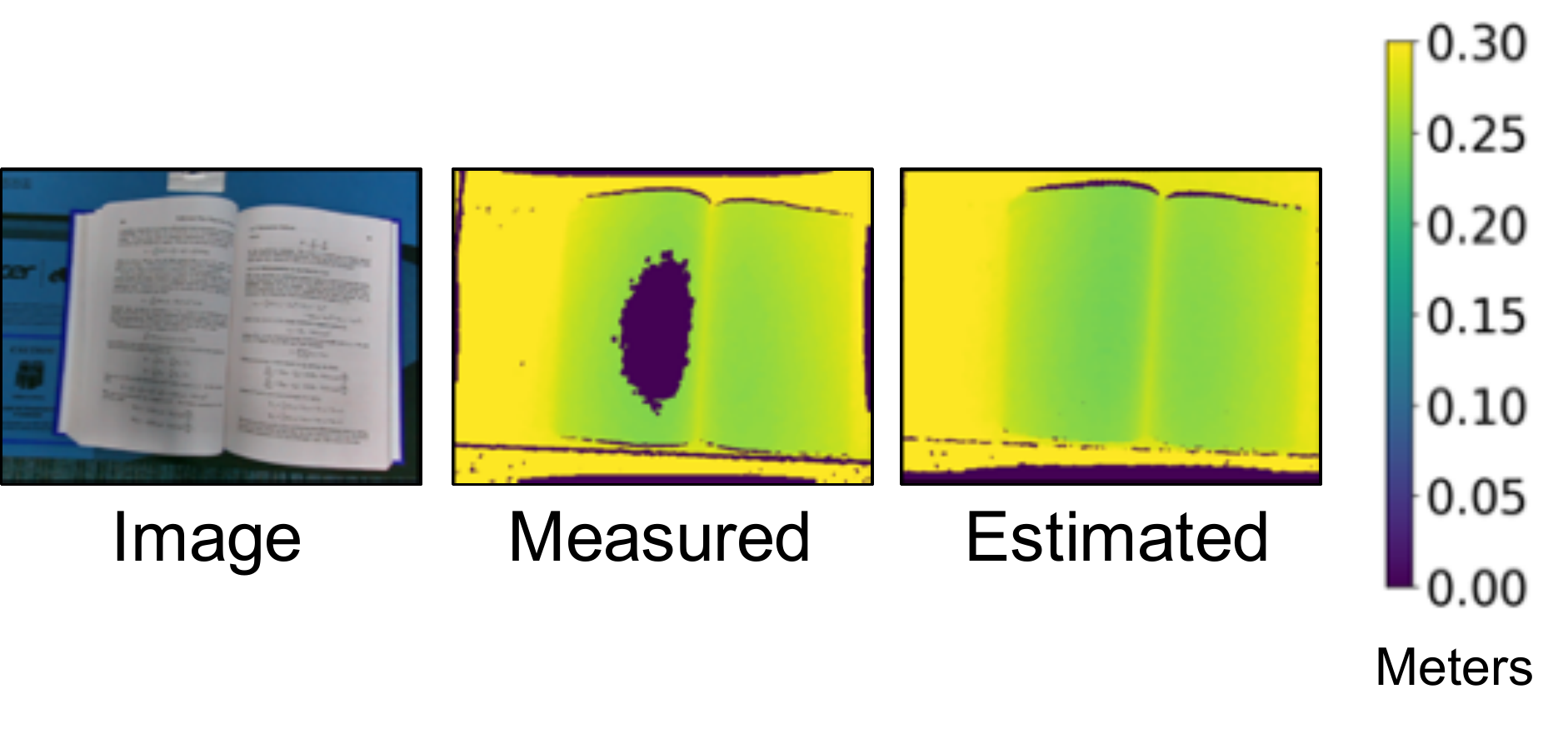}
	\caption{ \textbf{Saturation}: We estimate the depth for pixels that are saturated using our algorithm. The purple regions cannot be sensed by the TOF camera.}
	\label{fig:saturation}
\end{figure}

We first consider the scenario where the TOF camera goes out of range by acquiring images and depth maps of a scene shown in the first image of Figure \ref{fig:outofrange}. We use the Pico Zense DCAM710 RGB-D sensor \cite{Picozenze}, which contains a TOF and digital camera that outputs $640\times480$ depth maps and $1080\times1920$ RGB images, respectively. We expect that as we move the sensor away from the objects in the scene, we will not be able to measure depth for every object. We show an instance of this in the second image of Figure \ref{fig:outofrange}, where the TOF camera goes out of range and the depth for the box is unknown. To infill the depth values for the box, we use a previously measured image and depth map pair, where depth is available for both objects, and the current image to estimate a new depth map, which is shown in the last image of Figure \ref{fig:outofrange}. One limitation of our approach is that we can only infill regions where we have previous depth, and in this case, we cannot estimate depth for the wall. To evaluate the accuracy of our depth map, we compute the mean relative error for the overlapping pixels between the measured and estimated depth map. Because the scene is rigid, which means that the relative distance between the box and chair does not change, we expect that this mean relative error is also representative of what we would obtain if the depth for the box is available in the measured depth map. In this example, we achieve a mean relative error of 0.87\%.   

We also consider the scenario where a TOF camera becomes saturated by acquiring images and depth maps of a scene, shown in the first image of Figure \ref{fig:saturation}, as we move the TOF camera closer to the book. As shown in the second image of Figure \ref{fig:saturation}, the sensor saturates and depth is not available in the center of the book. By using a previous image and depth map pair, we are able to overcome this deficiency and estimate depth in this region, achieving a mean relative error of 0.6\% for the overlapping pixels. 

\section{Conclusion}
\label{sec:conclusion}

In this paper, we present an algorithm to estimate causal depth maps using concurrently collected images and previously measured depth. We use this approach to reduce the power of TOF imaging. Instead of using the TOF camera  continuously to acquire depth, we estimate depth maps using our technique and only use the TOF camera when an accurate depth map cannot be estimated. To ensure that the power for depth sensing is reduced, we design our algorithm to run efficiently on a low power embedded platform, carefully balancing the estimation of optical flow with that of pose. The resulting implementation produces $640\times480$ depth maps in real time, or 30 frames per second. We evaluated our approach on  several RGB-D datasets, where our technique produces depth maps with a mean relative error of 0.96\% and lowers the usage of the TOF camera by 85\%. When used with commercial TOF cameras, our algorithm can reduce the total power for depth sensing by up to 73\%. 

\section{Acknowledgement}

We thank Analog Devices for funding this research. We also thank the research scientists within the company for helpful discussions and feedback. 

\bibliographystyle{IEEEtran}

\end{document}